\documentclass[aps,preprint,showpacs,preprintnumbers,amsmath,amssymb, lettersize]{revtex4}

\usepackage{graphicx}

\begin{document}

\title{Algorithm for anisotropic diffusion in hydrogen-bonded networks.}
\author{Edoardo~Milotti}
\affiliation{Dipartimento di Fisica, Universit\`{a} di Trieste,
and INFN -- Sezione di Trieste, Via Valerio, 2, I-34127 Trieste, Italy}
\email{milotti@ts.infn.it}

\date{\today}

\begin{abstract}
In this paper I describe a specialized algorithm for anisotropic diffusion determined by a field of transition rates. The algorithm can be used to describe some interesting forms of diffusion that occur in the study of proton motion in a network of hydrogen bonds. The algorithm produces data that require a nonstandard method of spectral analysis which is also developed here. Finally, I apply the algorithm to a simple specific example.
\end{abstract}

\maketitle


\section{Introduction}
\label{intro}


Protons migrating in water have an anomalously high mobility \cite{agmon} and their diffusion is actually limited by the continuous rearrangement of hydrogen bonds \cite{agmon,agmon2}. Indeed protons migrating in ice move faster than protons in water, as the transition rate from one water molecule to the next is enhanced by the higher molecular order in ice. Proton mobility increases whenever water molecules are constrained, as in carbon nanotubes \cite{mh,mashi}. Local electric fields also orientate water molecules, and thus should lead to a local increase of proton mobility, and indeed it is now known that there is a definite water dipole orientational order in the hydration water close to ionizable residues in hydrated proteins \cite{higo,higo2,yoko,kumar}: this is a collective property, which is somewhat independent of the individual fluctuations of the water dipoles.

Here I am not concerned with the detailed simulation of proton motion which is the subject of several specialized papers like \cite{hal,cho,voth}, but rather I wish to set up the framework for a simulation of the random walks performed by protons in some interesting context, like proton migration on the surface of hydrated proteins \cite{carmil}. The basic idea is that protons move faster in the network of hydrogen bonds just where there is a higher molecular order, i.e., the transition rate is higher where there is higher spatial order, and because of the continuous rearrangement of the water molecules which make up a fluctuating bond structure, the random walk performed by a single proton can actually be viewed as a walk in continuous space and continuous time, as long as the time resolution of the process is longer than the relaxation time of water dipole motion. 

Here I take for granted that there is some induced order in the hydrogen bond network, like the dipole field described in \cite{higo,higo2}, and I introduce a corresponding field of transition rates $\gamma(\mathbf{r},t)$, such that the time-dependent probability density $\rho(\mathbf{r},t)$ and the associated probability $\Delta p = \rho \Delta V$ of finding a random walker (a proton) in the small volume $\Delta V$ at position $\mathbf{r}$ and time $t$, yield the following equation for the decrease of $\Delta p$, due to random walker escape from the region, 
\begin{equation}
\left. \frac{d}{dt}\Delta p(\mathbf{r},t)\right|_- = - \gamma(\mathbf{r},t)\Delta p(\mathbf{r},t)
\end{equation}
I also assume that $\gamma(\mathbf{r},t)$ is a continuous, differentiable function. 

\begin{figure}
\includegraphics[width=3.2in]{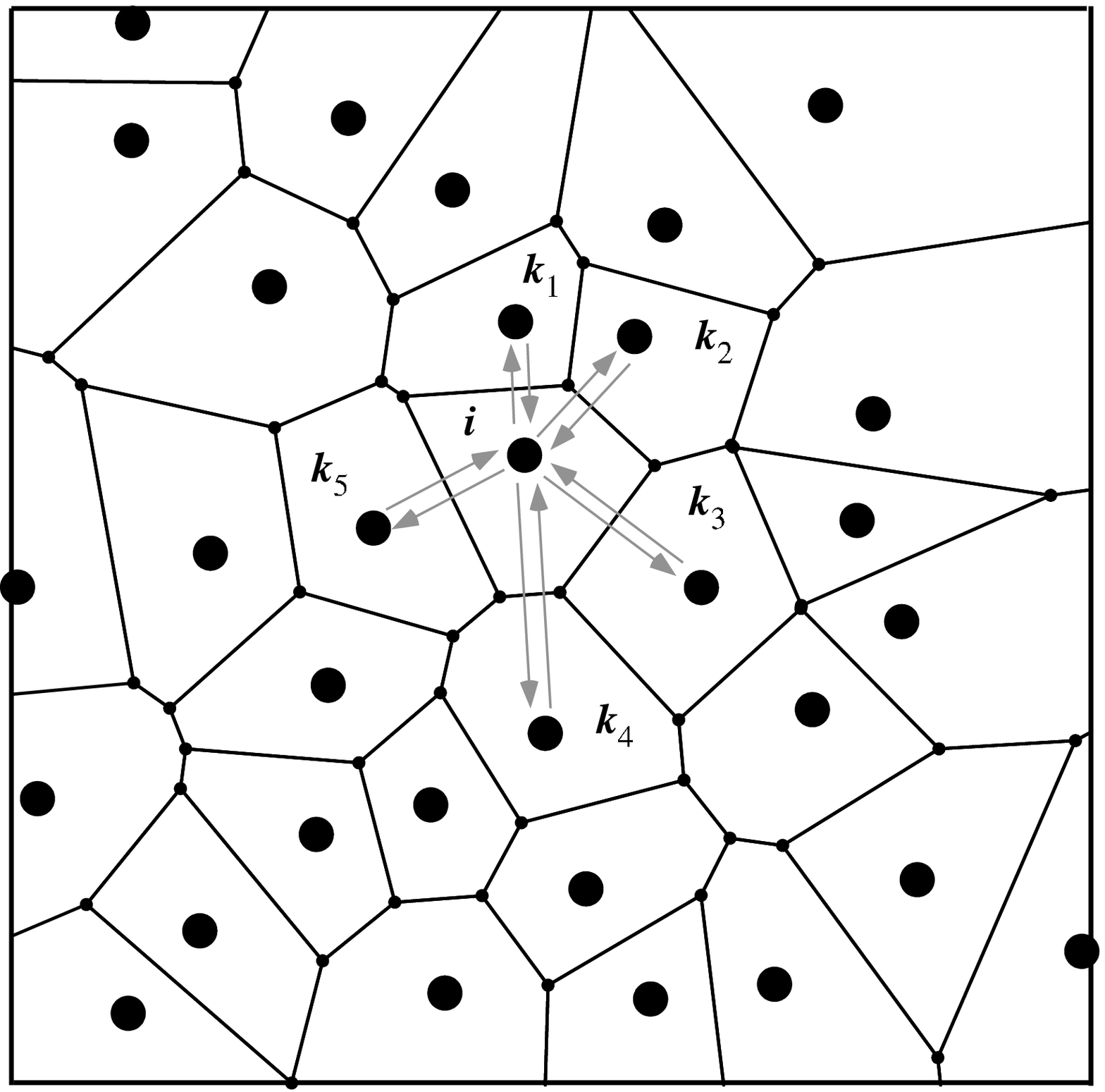}
\caption{\label{fig1} Here space has been subdivided in elementary regions (the dots represent the centroids of the elementary regions): each region is characterized by a particular transition rate $\gamma_i$ and by an elementary volume $V_i$. In this figure the transitions between the central region $i$ and the adjacent regions (denoted here by an index $k_j$) are marked by the gray arrows.}
\end{figure}

The situation is illustrated in figure \ref{fig1}, which shows a random subdivision of a plane region: a set of positions -- marked by the large black dots -- is associated to small surrounding regions; the arrows in the figure mark the flow of random walkers in the central region to and from the bordering regions. If the area (actually, the line length in this 2D representation) of the interface between the central region and the $k_j$-th region is $A_{i,k_j}$, and the total interface area of the $k_j$-th  region is $A_{k_j}$, then it is easy to see that the total derivative of $\Delta p_i$ is 
\begin{equation}
\frac{d\Delta p_i}{dt} = - \gamma_i\Delta p_{i} + \sum_j \frac{A_{i,k_j}}{A_{k_j}}\gamma_{k_j}\Delta p_{k_j}
\end{equation}
and the global flow in this discretized system is described by a system of coupled linear differential equations.

More importantly, we can define currents for the inflow and outflow of random walkers from a modified form of Fick's law
\begin{equation}
\label{current}
J_{i \rightarrow k} = \alpha \frac{\gamma_i \rho_i - \gamma_k \rho_k}{\Delta_{i,k}}
\end{equation}
where $\Delta_{i,k}$ denotes the distance between the centroids of the bordering $i$-th and $k$-th region, and the parameter $\alpha$ is akin to the diffusion coefficient, but is measured in different units (it has the dimensions of a surface) \cite{note1}. The current in the previous formula is actually a projection along the direction that connects the centroids of the bordering region and it is easy to generalize to the continuum case and find the outflowing current
\begin{equation}
\label{currentd}
\mathbf{J} = - \alpha \nabla(\gamma \rho)
\end{equation}
so that finally one finds the following Fokker-Planck equation from the conservation of the total number of random walkers
\begin{equation}
\label{fokplanck}
\frac{\partial \rho}{\partial t} = \alpha \nabla^2(\gamma \rho)
\end{equation}
assuming that $\alpha$ does not depend on position. 

In the following sections I describe an algorithm to simulate this kind of diffusive motion: first I discuss the angular distribution, then confinement to motion on surfaces,  and in section \ref{asynch} I show how to extend the algorithm for asynchronous updates. In section \ref{fft} I give a recipe to analyze asynchronous data. In section \ref{center} I discuss a simple example, and finally in section \ref{concl} I give a short summary and outlook for the utilization of the algorithm.

\section{Angular distribution}
\label{algo}

From equation (\ref{currentd}) we see that the current actually contains two contributions
\begin{equation}
\label{diffcurr}
\mathbf{J} = - \alpha \nabla(\gamma \rho) = - \alpha \left( \rho \nabla \gamma + \gamma \nabla \rho) \right)
\end{equation}
however when we consider the problem at hand -- namely, the diffusion of protons in the network of hydrogen bonds, and we remark that we wish to describe the individual proton motion, then we notice that we are only interested in situations where $\nabla \rho = 0$. In fact, protons repel other protons that are too close, and obey a sort of effective exclusion principle -- which is actually independent of their fermionic nature -- and the position of the individual proton corresponds to a peak of the instantaneous probability density: therefore the current defined in (\ref{currentd}) has the same direction as $\nabla \gamma$ in all cases of practical interest. This direction corresponds  to the average proton motion, but for a single transition to a nearby site it can only define the axis of an angular probability distribution. Here I make the simplest possible choice, namely that the angular probability distribution is a simple dipole distribution defined by the normalized conditional probability density for the unit vector $\mathbf{n}$
\begin{equation}
P(\mathbf{n}|\mathbf{n}_0) = \frac{1}{I_0} \left( 1 + \Delta P \mathbf{n}\cdot \mathbf{n}_0 \right)
\end{equation}
where $I_0$ is the normalization factor ($I_0 = 2\pi$ in the 2D case and $I_0 = 4\pi$ in the 3D case), and  $\mathbf{n}_0$ is the unit vector
\begin{equation}
\mathbf{n}_0(\mathbf{r}) = \frac{\nabla \gamma}{|\nabla \gamma|}
\end{equation}
so that the decrease of the density $\rho$ due to  the flow in the angular range $\Delta \Omega$, during a given time interval $\Delta t$, is 
\begin{equation}
\frac{\gamma \rho}{I_0} \left( 1 + \Delta P \mathbf{n}\cdot \mathbf{n}_0 \right) \Delta t \Delta \Omega
\end{equation}
The constant inside the parenthesis corresponds to the isotropic loss term, while the other term is associated to the current (\ref{diffcurr}). From a comparison of the elementary flows of random walkers in direction $\mathbf{n}$ we find 
\begin{equation}
\mathbf{J}\cdot \mathbf{n} = \gamma \rho \frac{\Delta P}{I_0} \mathbf{n}\cdot \mathbf{n}_0
\end{equation}
so that 
\begin{equation} 
\Delta P = I_0 \alpha \frac{|\nabla \gamma |}{\gamma}
\end{equation}
and the conditional angular probability density is
\begin{equation}
\label{ang}
P(\mathbf{n}|\mathbf{n}_0) = \frac{1}{I_0} \left( 1 +  I_0 \alpha \frac{|\nabla \gamma |}{\gamma} \mathbf{n}\cdot \mathbf{n}_0 \right)
\end{equation}
The conditional angular probability density (\ref{ang}) can be used to generate random walks discarding the time information. Here I take the following time-independent expression for the transition rate 
\begin{equation}
\label{gamma1}
\gamma(\mathbf{r},t) = \gamma(\mathbf{r}) =\frac{A}{|\mathbf{r}|^2 + \Gamma^2} + B
\end{equation}
which has an obvious symmetry center, located in the origin, which corresponds to the position of the peak value as well. This transition rate is motivated by the considerations put forward in the introduction: if protons migrate in a hydrogen bonded network with polarization centers that create partial ice-like order in their neighborhood, then the transition rate (\ref{gamma1})  is highest, and saturates, close to the polarization centers, and decays to a constant value with a $1/r^2$ behavior for $r\gg \Gamma$ (i.e., it has a radial dependence like the potential of electric dipole fields). Notice also that the anisotropy coefficient is
\begin{equation}
\frac{|\nabla \gamma |}{\gamma} = \frac{2|\mathbf{r}|}{|\mathbf{r}|^2 + \Gamma^2}
\end{equation}
The techniques to generate random angles which are distributed according to the probability density (\ref{ang}) are reviewed in appendix \ref{appA}, and figures \ref{fig2}-\ref{fig5} show some examples: in these examples all length and distance units are in arbitrary units. Figure \ref{fig2} shows random walks around a single center with transition rate (\ref{gamma1}): the random walker starts at the origin, with a fixed step length $ = 0.005$ arbitrary units; the horizontal and vertical scales are also labeled with the same arbitrary length units; the parameters of the transition rate function are the same in these simulations, $A=1$, $B=0.1$, and $\Gamma = 1$, while $\alpha$ changes in the three cases displayed in the figure. Larger values of $\alpha$ correspond to higher anisotropy, and we see that as the anisotropy grows, the random walk becomes more and more compact.

Figure \ref{fig3} shows a random walk with two centers at positions $\mathbf{r}_1=(-1,0)$, $\mathbf{r}_2=(1,0)$ (arbitrary units): the random walker starts at the origin, with a fixed step length $ = 0.01$ arbitrary units; the horizontal and vertical scales are also labeled with the same arbitrary length units. In this case the transition rate is similar to (\ref{gamma1}), but with two centers,
\begin{equation}
\label{gamma2}
\gamma(\mathbf{r})=\frac{A_1}{|\mathbf{r}-\mathbf{r}_1|^2 + \Gamma_1^2} + \frac{A_2}{|\mathbf{r}-\mathbf{r}_2|^2 + \Gamma_2^2} + B
\end{equation}
with $A_1=A_2=1$, $B=0.1$, and $\Gamma_1=\Gamma_2=1$, and $\alpha = 1$. Here the random walker explores the regions around both centers. 

Figure \ref{fig4} shows a situation which is similar to figure \ref{fig3}, although it is more complex. The transition rate is once again similar to (\ref{gamma1}), but now it has ten centers, 
\begin{equation}
\label{gamma10}
 \gamma(\mathbf{r}) = \sum_{k=1}^{10} \frac{A_k}{|\mathbf{r}-\mathbf{r}_k|^2 + \Gamma_k^2} + B
\end{equation}
with $A_k=1$, $B=0.1$, and $\Gamma_k=0.1$, and $\alpha = 0.025$; the step length is $ = 0.01$ arbitrary units. The centers are scattered randomly, with a lower bound on the minimum distance between them; the figure shows three snapshots at different times in the simulation, as the random walker starts from the center of the figure, drifts to one of the centers and later migrates to other neighboring centers. 

Finally figure \ref{fig5} shows a random walk in space about two centers at $\mathbf{r}_1=(-1,0,0)$, $\mathbf{r}_2=(1,0,0)$ (arbitrary units), which is very similar to the random walk in figure \ref{fig3}: the transition rate is still given by expression (\ref{gamma2}), with $A_1=A_2=1$, $B=0.1$, and $\Gamma_1=\Gamma_2=1$, and $\alpha = 0.5$, with a fixed step length $ = 0.1$ arbitrary units. Once again the random walker explores the regions around both centers.

\section{Diffusion on surfaces}
\label{confine}

In many cases it is important to confine the motion of the random walkers to some particular portion of space, for instance in the case of protons on hydrated proteins the motion is confined to the thin hydration layer. The simulation method outlined in the previous section can be adapted to provide such a confinement to a surface: in this case one can define at each step the tangent plane at the position of the random walker and proceed as in the 2D case. Obviously the gradient of the transition rate $\gamma$ used in the formulas of the previous section must be projected on the tangent plane, and moreover the directions must be generated according to the 2D angular distribution (see appendix \ref{appB}). 
Figure \ref{fig6} shows a random walk on a spherical surface with 10 centers as in the examples in the previous section: the sphere has radius 1 (arb. units), the transition rate is given by expression (\ref{gamma10}), with $A_k=10$, $B=1$, and $\Gamma_k=0.25$, and $\alpha = 0.2$; the step length is $ = 0.1$ arbitrary units.

\section{Asynchronous transitions}
\label{asynch}

In the previous sections we have discussed the space behavior of the random walks, but obviously we can use the transition rate function $\gamma(\mathbf{r},t)$ to describe the time behavior as well. It is possible to choose a fixed time step $\Delta t$ and use the transition rates to compute the probability of generating a transition in the time interval $(t, t + \Delta t)$ (synchronous update), however this is inconvenient if the function $\gamma(\mathbf{r},t)$ spans a wide range of values, because it means that the choice $\Delta t \ll \min (1/\gamma(\mathbf{r},t))$ which is required for an accurate simulation, produces very long waiting times where the transition rate is very small (and therefore, very large amounts of sampled data). It is actually much more practical to use the transition rate to  generate directly  the transition times of each step, which we assume to be independent (asynchronous update) from one another. With this -- rather natural -- assumption of independency, it is very simple to generate the transition times, as explained in appendix \ref{appB}, although this leads to uneven sampling, and requires a specialized form of spectral analysis. 

\section{Fourier analysis of asynchronous data}
\label{fft}

Using asynchronous sampling times it is not possible to use the standard Fourier or other similar spectral analysis techniques \cite{km}. However the signal produced by the time-domain simulation is ``exact'', at least in the sense that there are no algorithmic artifacts due to sampling and it is desirable to extract as much information as possible. To this end, I notice that any function of the position of the random walkers must be stationary between successive transitions, and that it is possible to make direct use of the definition of Fourier transform
\begin{equation}
F(\omega) = \int_{-\infty}^{+\infty} s(t) e^{-i \omega t} dt
\end{equation}
where $s(t)$ is any signal produced in the simulation, which depends on the positions of the random walkers (e.g., a component of the electric dipole moment if the random walkers are charged particles). The signal $s(t)$ has the fixed value $s_n$ in the time interval $(t_n, t_n+1)$, where $t_n$ is the time of the $n$-th transition, and we find: 
\begin{eqnarray}
\nonumber
F(\omega) & = & \sum_n s_n \int_{t_n}^{t_{n+1}} e^{-i \omega t} dt \\
\label{sinc}
& = & \sum_n s_n e^{-i \omega (t_{n+1} + t_n)/2} (t_{n+1} - t_n) \frac{\sin\left[ \omega (t_{n+1} - t_n)/2\right]}{\omega (t_{n+1} - t_n)/2} \\
\nonumber
& = &\frac{2}{\omega} \sum_n s_n \sin\frac{\omega (t_{n+1} - t_n)}{2} \cos \frac{\omega (t_{n+1} + t_n)}{2} \\
\label{fourier}
&& - \frac{2i}{\omega} \sum_n s_n \sin\frac{\omega (t_{n+1} - t_n)}{2} \sin \frac{\omega (t_{n+1} + t_n)}{2}
\end{eqnarray}
Using equation (\ref{fourier}) the Fourier transform can be evaluated exactly for all frequencies, and without aliasing: in practice this is possible, practical, and actually useful only for a small finite set of frequencies. If we had used a Discrete Fourier Transform (DFT) algorithm \cite{km} to analyze $N$ real samples, we would have found $N/2$ independent Fourier coefficients, and using a Fast Fourier Transform (FFT) algorithm the time complexity of the calculation would be $O(N \log N)$. If we use the algorithm defined by equation (\ref{fourier}) to compute $M$ values ($M \le N/2$) of the Fourier transform, the time complexity is clearly $O(NM)$, so that in a practical calculation we can only compute a reduced number of Fourier coefficients. However, I remark that in addition to being exact, the algorithm has another major advantage over the standard DFT calculations: there is no limitation to the set of frequencies that can be computed, and in particular one can choose a set of frequencies that is not evenly spaced and that is denser close to the origin, which is particularly useful in this case since the random walk -- when considered as a noise process -- is expected to produce a spectrum with a large power-law peak at low frequencies.

This kind of analysis is actually limited by the finite time span of the generated signal: we see from equation (\ref{sinc}) that the Fourier transform of the generated signal is a sum of sinc functions, and therefore it is not useful to represent the transform for frequencies lower than $\omega_{min} = \pi/T$ where $T$ is the signal duration and $\omega_{min}$ is the lowest positive zero of the corresponding sinc function. With this limitation, we can sample the Fourier transform at frequency values that are evenly spaced on a logarithmic scale and obtain a better representation of the transform close to the origin than is possible with conventional methods.

In this approach we evaluate the Fourier transform of the simulated signal in the time interval $(0,t_N)$ and we implicitly assume that the signal vanishes outside this interval: this is different from the standard (implicit) assumption in standard DFT analysis, where the observed signal is assumed to repeat periodically outside the observation interval \cite{km}. If we introduce a rectangular window with a width equal to the observation interval ($t_N$), we see that the present method returns a Fourier transform that is the convolution of the transform of the signal with the transform of the rectangular window, which is
\begin{equation}
\label{rect}
\int_0^{t_N} \exp(-i\omega t) dt = t_N \exp(-i \omega t_N/2) \frac{\sin(\omega t_N/2)}{(\omega t_N/2)}
\end{equation}
As a consequence of the convolution associated to the rectangular window we see that a constant nonzero level produces a sharp peak centered at zero frequency, with a shape given by equation (\ref{rect}); this peak corresponds to the standard DC peak in DFT analysis, and has tails with a $1/\omega^2$ spectral behavior that may mimic the low-frequency behavior of  a standard Debye relaxation with a very small decay rate. The mean level of the simulated signal is 
\begin{equation}
\bar{s}=\frac{1}{t_N} \sum_n s_n \int_{t_n}^{t_{n+1}} dt = \frac{1}{t_N} \sum_n s_n \left(t_{n+1} - t_n \right)
\end{equation}
and thus we can correct for the DC peak by subtracting its transform 
\begin{equation}
\frac{2\bar{s}}{\omega}  \exp(-i \omega t_N/2) \sin(\omega t_N/2)
\end{equation}
from the signal transform.

\section{Random walk about a single center}
\label{center}

As an example, I consider here a complete simulation (3D space and time data) for random walks about a single center (as defined by the transition rate (\ref{gamma1}) ) located in the origin. 
The transition rate function used in this example is given again by expression (\ref{gamma1}), with $A=100$, $B=10^{-8}$, $\Gamma = 0.1$, $\alpha=0.001$, and with a step length $= 0.01$. The values of $A$ and $B$ have been chosen to maximize the range of $\gamma$ sampled by the random walker, while still keeping a rather short simulation time. I have generated 500 random walks and 10000 transitions for each walk; the random walker always starts at the origin (where the center of (\ref{gamma1}) is located). The results of the simulation are shown in figures \ref{fig7}-\ref{fig11}: figure \ref{fig7} is the superposition of a few walks, and it is qualitatively clear that the density profile is very similar to that of the standard random walk in the plane. Figure \ref{fig8} shows instead the $x$ projection of the position signal vs. time, and this is not very different, e.g., from an electric dipole component if the random walkers are charged particles. 
The insets show parts of the signal with increasing magnification, and the last inset displays clearly the stationary parts of the signal between successive transitions.
Figure \ref{fig9} shows the (unnormalized) distribution of time intervals between successive transitions: the figure demonstrates clearly that although the times have been generated according to the interval distribution in appendix \ref{appB}, the distribution in figure \ref{fig9} is not a simple exponential, but rather it contains two different power-law regions (marked by the dotted lines in this log-log plot), which reflects the way in which the transition probability function is sampled by the random walks.
One well-known property of ordinary random walks is that their mean square radius is proportional to time, i.e., $\langle r^2 \rangle \propto t$: here we see (figure \ref{fig10}) that this linearity is recovered only asymptotically, as random walkers explore regions that are far away from the origin. 
Finally, I have used the power spectral estimation method of section \ref{fft} to analyze the $x$ position signals (as those in figure \ref{fig8}): the result is shown in figure \ref{fig11}. Figure \ref{fig11}a is the power spectrum obtained in a single realization of the random walk, while figure \ref{fig11}b is the average of 400 spectra. In each part of figure \ref{fig11}, the thin gray line represents an ideal power-law spectral density with the same slope as the the average of 400 spectra, i.e., a $1/f^2$ spectrum, which is usually expected in these types of processes. In fact the random walks simulated here effectively sample asymptotically only a rather limited range of transition rates -- even though the $A/B$ ratio is quite high --  and this means that the usual superposition argument that leads to more general power-law spectra \cite{weiss} does not apply here and it is quite natural to find a $1/f^2$ spectrum.

\section{Discussion}
\label{concl}

Before concluding this paper it is important to note that the correct continuum formulation of the diffusion equation in an inhomogeneous environment has been the subject of much discussion in the past and  is still debated (see \cite{vankam,chrisped}, and references therein). The generalization is unclear because the microscopic details seem to matter \cite{vankam}. Moreover, there is also some interest towards the diffusion equation in various forms of anomalous diffusion \cite{sb}. Here I wish to stress again that the results presented in this paper are specialized and are meant to address diffusion in structures like those described in \cite{higo,higo2}, unlike other approaches described in the existing literature that deal with more general diffusion problems \cite{chris,kiku1,kiku2}; still, the diffusion equation (\ref{fokplanck}) is similar to equation (5) in reference \cite{vankam}, and therefore it is interesting to give one further look at its structure. In section \ref{algo} we have seen that the current (\ref{currentd}) has two components, and the gradient term that generates the random walks discussed here roughly corresponds to the so-called ``spurious'' drift term (using the terminology of  reference \cite{vankam}). The other term in the current, namely $- \alpha \gamma \nabla \rho$ has been neglected because the single random walkers considered here are charged fermions and obey a sort of effective exclusion principle. Indeed, in a context like that of \cite{higo,higo2} protons repel because of their charge, while their spin structure, and therefore also their true fermionic character -- and the Pauli exclusion principle -- do not matter much; this effective exclusion principle has been used in the past for an Ising-like modeling of proton motion, where the presence or absence of a proton at a given position is treated like a pseudo-spin variable (see the discussion in the review paper \cite{car}). The situation would be very different if space could be filled by a cloud of random walkers: in a case such as this -- which roughly corresponds to random walkers that effectively behave as bosons -- the neglected term should be included. Thus the actual importance of the different terms of the diffusion equation (\ref{fokplanck}) depends on the bosonic or fermionic character of the random walkers.

One prominently missing term in the diffusion equation (\ref{fokplanck}) is the usual drift term associated to external fields, however if we look at the structure of the current (\ref{current}), we see that we can easily produce a flow unbalance with a space- (and possibly time-) dependent $\alpha$, so that we obtain a modified diffusion equation
\begin{equation}
\label{fokplanck2}
\frac{\partial \rho}{\partial t} =  \nabla \cdot \left[ \alpha \nabla(\gamma \rho) \right] = \nabla \alpha \cdot  \nabla(\gamma \rho) + \alpha  \nabla^2(\gamma \rho) 
\end{equation}
with an additional $\alpha$-dependent drift term.

A final, important comment, is that the algorithm presented here has a sort of backward approach with respect to other existing algorithms for random walks and diffusion in inhomogeneous environments, as it starts directly from the transition rates, instead of deriving them from a given diffusion equation (as, e.g., \cite{chris}).

To conclude, in this paper I have described a novel algorithm for anisotropic diffusion, which is continuous both in space and time, and I have discussed its application to a simple example, in anticipation of further work that shall be carried out in a realistic simulation of noise in proton conduction in weakly hydrated proteins \cite{carmil2}.

\appendix
\section{Angular distributions}
\label{appA}

Here I consider first the planar angular distribution defined by the normalized probability density
\begin{equation}
\label{dens2D}
P(\varphi;g) = \frac{1}{2\pi}(1 + g \cos \varphi)
\end{equation}
Using the standard inversion method (described in many textbooks, see, e.g., \cite{NR}), one finds that the solution $\varphi$ of the nonlinear equation 
\begin{equation}
\frac{1}{2\pi}(\varphi + g \sin \varphi)+\frac{1}{2} = y
\end{equation}
has the distribution described by (\ref{dens2D}) if $y$ is a uniform variate on the $(0,1)$ interval. 

The generation of a random direction in space from the normalized probability density 
\begin{equation}
P(\mathbf{n}; g) = \frac{1}{4\pi} (1 + g \mathbf{n}\cdot \mathbf{n}_0)
\end{equation}
requires two angles, a zenithal angle $\theta$ and azimuthal angle $\varphi$, where $\mathbf{n}_0$ defines the zenithal axis, so that the probability of finding a unit vector $\mathbf{n}$ in the interval $(\theta, \theta+d\theta)$, and $(\varphi, \varphi+d\varphi)$ is 
\begin{equation}
dP(\theta, \varphi; g) =\left[ \frac{1}{2} \left(1 + g \cos \theta\right) d \cos\theta \right] \left(  \frac{1}{2\pi} d\varphi \right) = \left[  \frac{1}{2} \left(1 + g x\right) d x \right] \left(  \frac{1}{2\pi} d\varphi \right)
\end{equation}
i.e., the probability density is the product of two independent densities, one uniform with respect to $\varphi$ on the $(0,2\pi)$ interval, and the other linear with respect to $x = \cos \theta$ on the $(-1,1)$ interval. Using again the inversion method, one finds that 
\begin{equation}
x = \frac{2}{g} \left[ -\frac{1}{2} + \sqrt{\frac{1}{4}+g\left( \frac{g}{4}-\frac{1}{2}+y \right)} \right]
\end{equation}
has the required linear distribution if $y$ is a uniform variate on the $(0,1)$ interval. 
Since $\mathbf{n}_0$ is not usually parallel to the $z$ direction, two rotations are also required: one first rotates the reference frame so that $\mathbf{n}_0$ is parallel to the $z$ axis, this is followed by the angle generation step, and finally one must transform back to the original reference frame.

\section{Time distribution}
\label{appB}

Time transitions are generated according to the exponential distribution, which has the probability density
\begin{equation}
\frac{dp_t}{dt} = \gamma \exp(- \gamma t)
\end{equation}
where $\gamma$ is the transition rate. The standard inversion method \cite{NR} can be used again to generate times 
\begin{equation}
t = - \frac{1}{\gamma} \ln y
\end{equation}
that are exponentially distributed if $y$ is a uniform variate on the $(0,1)$ interval.

\acknowledgments{I wish to thank warmly Giorgio Careri for his insightful comments and suggestions: he was the first to pinpoint the problem that spurred the research described here, and this paper would not have been written without his encouragement.

I also wish to thank Alessio Del Fabbro for his careful reading of the manuscript and for several interesting discussions.}

\pagebreak


\begin{figure}
\includegraphics[width=2.4in]{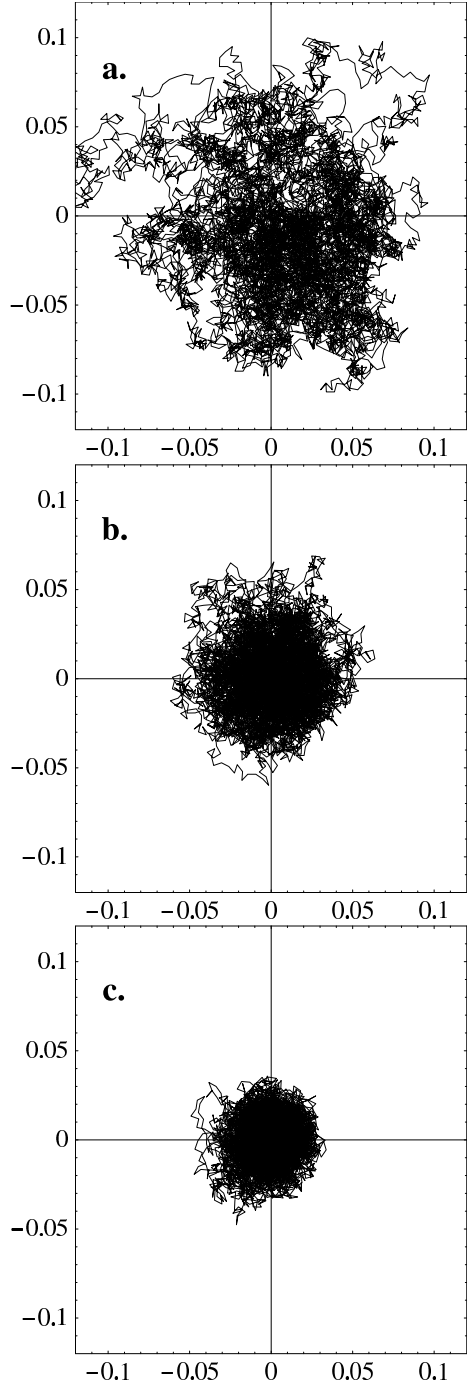}
\caption{\label{fig2} Random walks around a single center as explained in section \ref{algo}. The random walker starts at the origin, with a fixed step length $ = 0.005$ arbitrary units; the horizontal and vertical scales are also labeled with the same arbitrary length units. The width parameter is $\Gamma = 1$ in all simulations, and the other parameters have fixed values as well, $A=1$, $B=0.1$, while {\bf a.} $\alpha = 1$; {\bf b.} $\alpha = \sqrt{10}$; {\bf c.} $\alpha = 10$. Larger values of $\alpha$ correspond to higher anisotropy, and we see here that as the anisotropy grows, the random walk becomes more and more compact.}
\end{figure}

\begin{figure}
\includegraphics[width=3.2in]{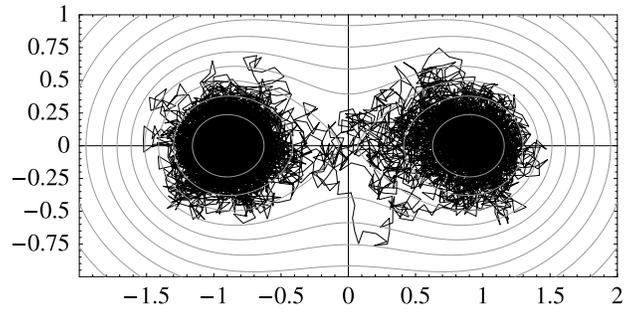}
\caption{\label{fig3} Random walk with two centers, as explained in the text. The walk is superposed on the contour plot of the transition rate $\gamma(\mathbf{r})$. The random walker starts at the origin, with a fixed step length $ = 0.01$ arbitrary units; the horizontal and vertical scales are also labeled with the same arbitrary length units. }
\end{figure}

\begin{figure}
\includegraphics[width=5in]{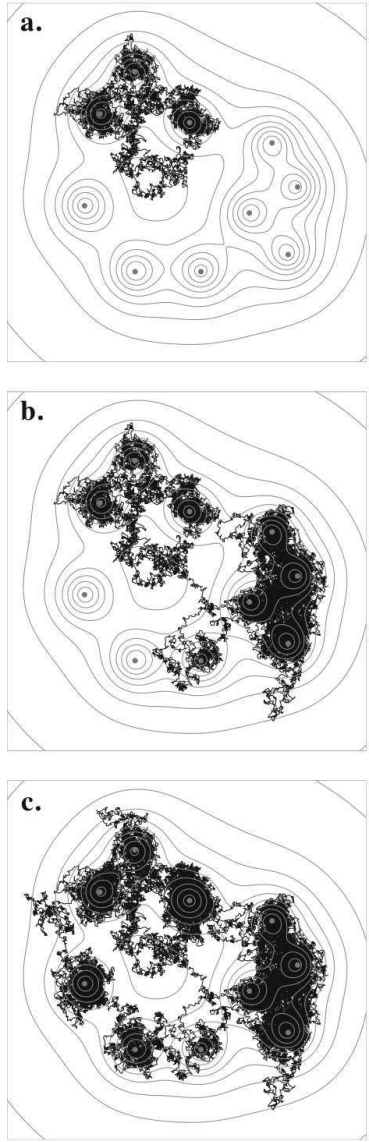}
\caption{\label{fig4} Random walk with several (10) centers, {\bf a.} after 10000 steps; {\bf b.} after 40000 steps; {\bf c.} after 70000 steps. The walk is superposed on the contour plot of the transition rate $\gamma(\mathbf{r})$.}
\end{figure}

\begin{figure}
\includegraphics[width=3.2in]{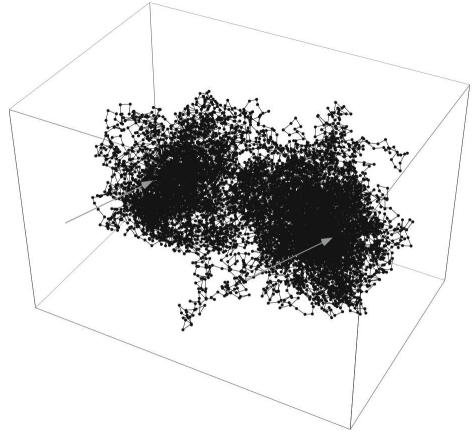}
\caption{\label{fig5} Random walk in space with two centers, located at the positions marked by the gray arrows.}
\end{figure}

\begin{figure}
\includegraphics[width=3.2in]{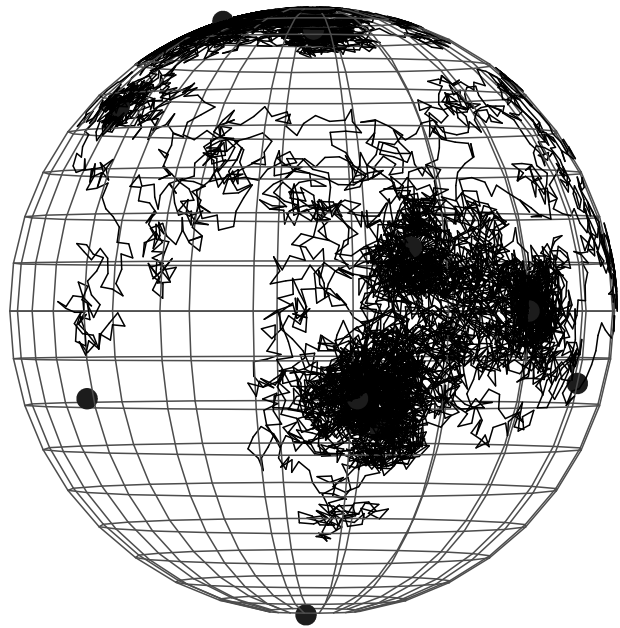}
\caption{\label{fig6} Random walk on a spherical surface, with several (10) centers, some of which have not yet been visited by the random walker.}
\end{figure}

\begin{figure}
\includegraphics[width=5in]{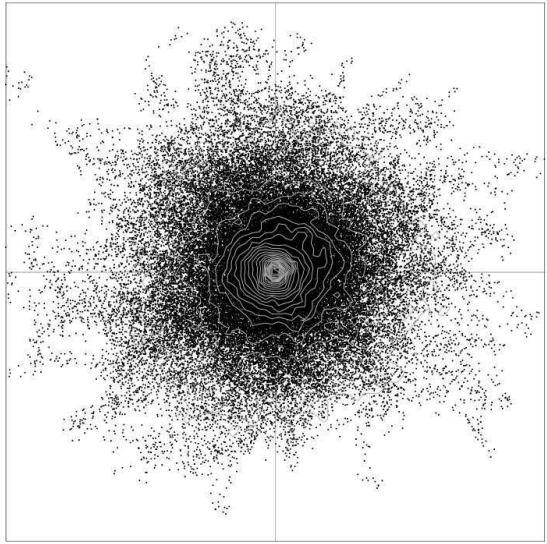}
\caption{\label{fig7} Superposition of several random walks about a single center in the origin. Isodensity contour lines are superposed on the density plot.}
\end{figure}

\begin{figure}
\includegraphics[width=3.2in]{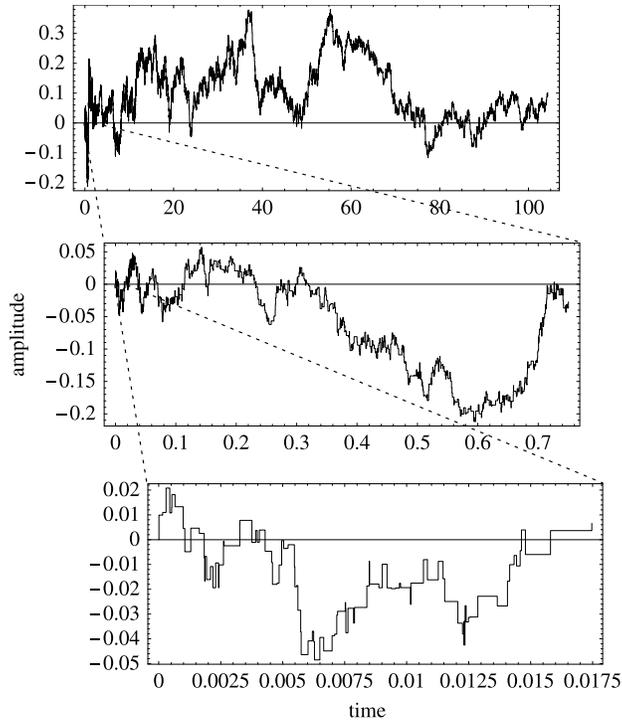}
\caption{\label{fig8} This figure shows the $x$ projection of the position signal (this is not very different from an electric dipole component if the random walkers are charged particles) vs. time for one of the random walks in the example of section \ref{center}; both position and time are in arbitrary units. The insets show parts of the signal with increasing magnification, and the last inset displays clearly the stationary parts of the signal between successive transitions.}
\end{figure}

\begin{figure}
\includegraphics[width=3.2in]{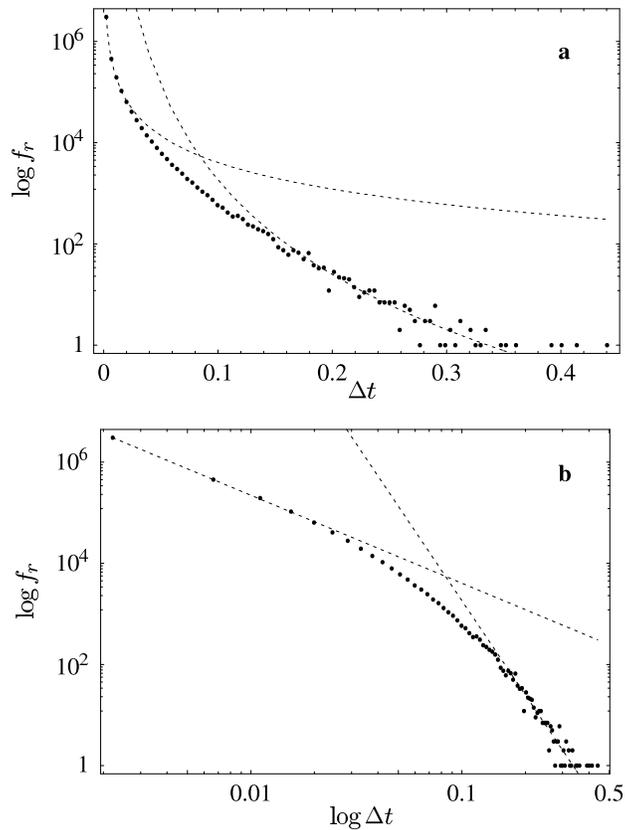}
\caption{\label{fig9} Unnormalized distribution of the time intervals $\Delta t$ between transitions in the set of 400 random walks of 10000 steps each described in the text: {\bf a}. logarithm of the relative frequency vs. $\Delta t$ (both in arbitrary units), which shows that the distribution is not a simple exponential, but rather contains two different power-law regions (dotted lines); {\bf b}. log-log plot of the same distribution, where the two power laws are identified by the nearly straight sections.}
\end{figure}

\begin{figure}
\includegraphics[width=3.2in]{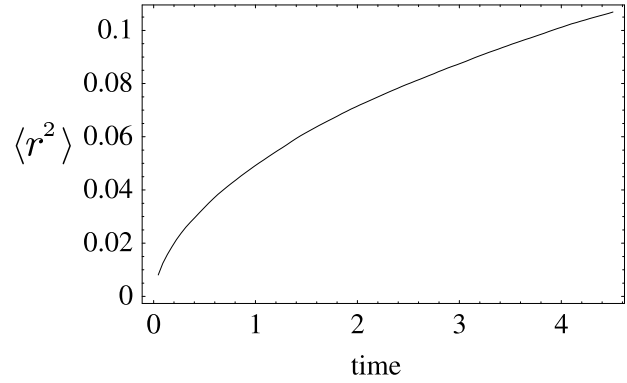}
\caption{\label{fig10} Mean squared distance $\langle r^2 \rangle$ vs. time. In an ordinary random walk the mean squared distance is a linear function of time: here we see that linearity is recovered only asymptotically, as the random walkers explore regions that are further away from the origin, where the $\gamma$ is nearly constant. }
\end{figure}

\begin{figure}
\includegraphics[width=3.2in]{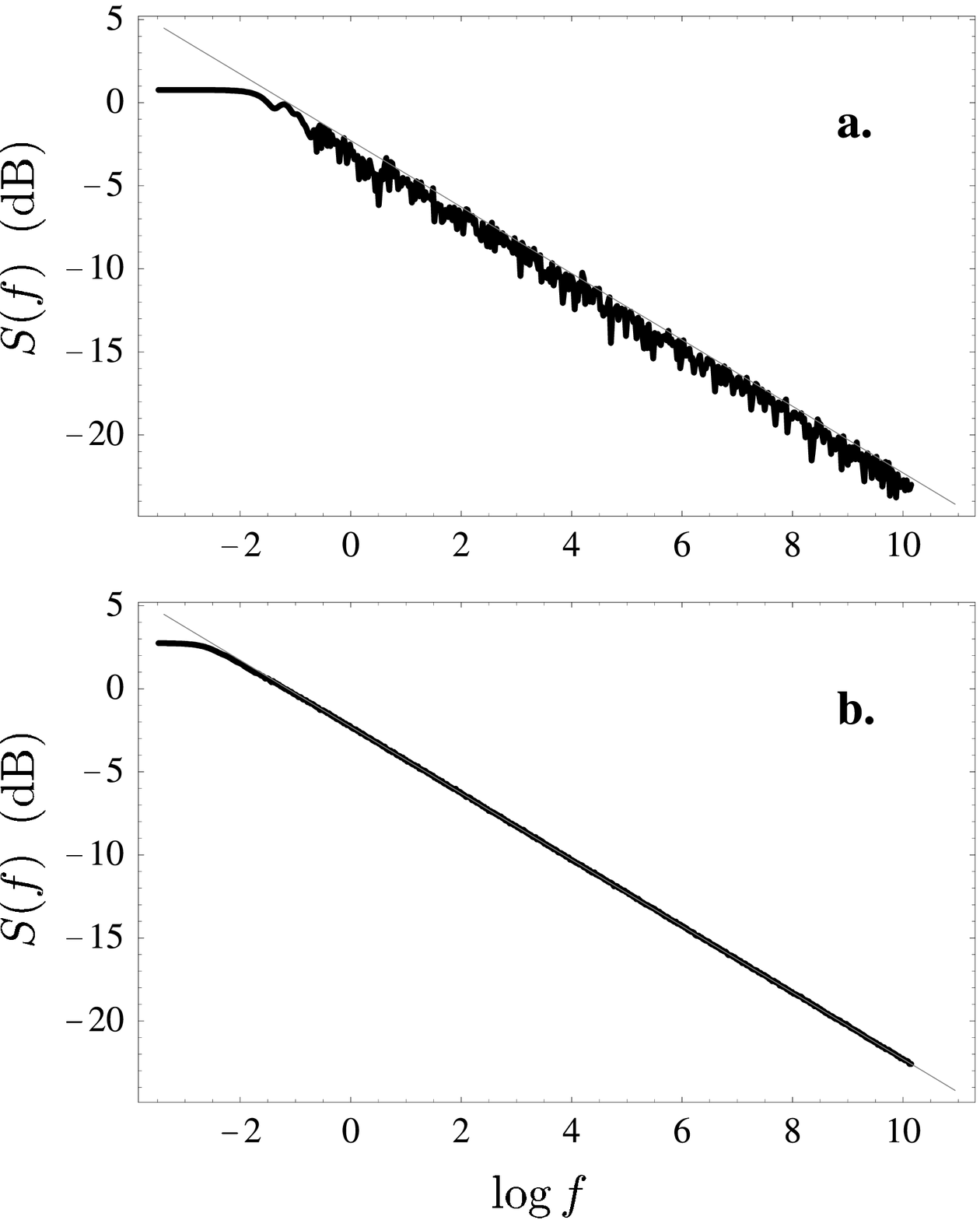}
\caption{\label{fig11} Spectral density calculated with the method of section \ref{fft}, sampled at logarithmically spaced frequencies: {\bf a.} spectrum obtained from the $x$ position signal for a single random walk; {\bf b.} average of 400 spectra. Spectral densities and frequencies are in arbitrary units. The thin gray line represents an ideal $1/f^2$ spectrum, which is expected for this kind of processes: the computed spectrum deviates from the ideal spectrum only at very low frequency, because of the limited observation time. Notice also that there is no upward bend at very high frequency -- a hint of the absence of aliasing.  In part {\bf a.} it is clearly visible that the spectrum is sampled at (500) logarithmically spaced frequencies, because there is no crowding at the high end of the spectrum, and no rarefaction at low frequency, unlike spectra obtained with the Fast Fourier Transform or other similar algorithms.}
\end{figure}


\begin{thebibliography}{00}
\bibitem{note1} The distance $\Delta_{i,k}$ in equation (\ref{current}) in the text is defined exactly only in the continuous limit, however this is not important because the discrete formulation in section \ref{intro} is used only to introduce the diffusion equation and is not relevant for the algorithm.
\end{thebibliography}
\end{document}